\documentclass[aps,prl,twocolumn,showpacs]{revtex4}

\usepackage{bm}
\usepackage{graphicx}
\usepackage{amsmath}
\usepackage{amssymb}
\usepackage[dvips]{color}
\usepackage{subfigure}

\begin{document}
\def\be{\begin{equation}}
\def\ee{\end{equation}}

\def\bfi{\begin{figure}}
\def\efi{\end{figure}}
\def\bea{\begin{eqnarray}}
\def\eea{\end{eqnarray}}

\title{Synchronized oscillations and acoustic fluidization in confined granular materials} 
\author{F. Giacco$^{1}$, L. de Arcangelis$^{2}$, M. Pica Ciamarra$^{3,4}$, E. 
Lippiello$^{1}$} 
\affiliation{%
$^{1}$Dept. of Mathematics and Physics, University of Campania ``L. Vanvitelli'', 
Caserta, Italy \\
$^{2}$Dept. of Industrial and Information Engineering,  University of Campania ``L. Vanvitelli'', Aversa (CE), Italy\\
$^{3}$Division of Physics and Applied Physics, School of
Physical and Mathematical Sciences, Nanyang Technological University, Singapore \\
$^{4}$CNR--SPIN, Dept. of Physics, University of Naples ``Federico II'', 
Naples, Italy
}


\date{\today}

\begin{abstract}
According to the acoustic fluidization hypothesis, 
elastic waves at a characteristic frequency form 
inside seismic faults even in absence of an external perturbation. These waves  are able to generate a normal stress which contrasts
the confining pressure and promotes failure.
Here we study  the mechanisms responsible for this wave activation via  numerical simulations of a granular fault model. 
We observe the particles belonging to the percolating backbone, which sustains the stress, to perform synchronized oscillations over elliptic-like trajectories in the fault plane. These oscillations occur at the characteristic frequency of acoustic fluidization. As the applied shear stress increases, these oscillations become perpendicular to the fault plane just before the system fail, opposing to the confining pressure, consistently with the acoustic fluidization scenario.
The same change of orientation can be induced by external perturbations at the acoustic fluidization frequency.
\end{abstract}


\maketitle

Confined granular materials under shear
 display the typical stick-slip dynamics observed in real fault systems. In the last years this dynamics has been deeply investigated in several experimental settings as well as by means of molecular dynamics simulations \cite{johnsonjia2005,johnsonsava2008,capozzaprl,petri_europjournal,KB11,johnson2012,
vanderelst2012,giaccopre,griffa2013,3ddegriffa2014,xia2011pre,xiamarone2013,johnson2016,santibanez2016}. These studies mostly focus on two central questions: i) Why the stress responsible for seismic failure is usually orders of magnitude smaller than the value expected on the basis of rock fracture mechanics?   
ii) Why seismic faults are very susceptible to even small amplitude transient seismic waves? 
Indeed, 
the resistance to shear stress of seismic faults is typically much larger 
than the one obtained in  experiments
measuring the friction coefficient of sliding rocks~\cite{hickman91}. Furthermore, remote triggering of earthquakes~\cite{hill93,gomberg2004,gomberg2005,GRL:GRL50813,LGV15} at distances of thousand kilometers from the main
shock epicenter indicates a high susceptibility of seismic faults to the passage of  seismic waves. 
 The hypothesis of Acoustic Fluidization (AF), formulated by Melosh \cite{melosh79,melosh96}, provides an answer to both questions. According to AF, the elastic waves produced by seismic fracture, at a characteristic frequency $\omega_{AF}$, diffuse and scatter inside the fault and then generate a normal stress which can contrast the confining pressure. In this way seismic failure is promoted. 
To investigate this scenario, experimental studies~\cite{johnsonjia2005,johnsonsava2008,johnson2012,vanderelst2012} have demonstrated that acoustic perturbations modify
granular rheology and lead to auto-acoustic compaction~\cite{vanderelst2012}. 
Recently the AF scenario has been explored in 3D molecular dynamics simulations \cite{giaccoprl2015} which have shown that weak external perturbations, at the frequency $\omega_{AF}$, even if increasing the confining pressure or reducing the applied shear, induce slip instabilities. Interestingly simulations have also shown that oscillations at the frequency $\omega_{AF}$
are activated immediately before each slip, even in the absence of an external perturbation. Nevertheless the mechanisms responsible for this activation,  as well as the non-linear response to external perturbation, are not  fully understood. 

In this article we shed light on these mechanisms by means of a detailed investigation of grain trajectories during the stick phase in 3D molecular dynamics simulations. Differently from previous studies, which mainly focus on the confining plate dynamics, we follow the evolution of each grain. This analysis shows that oscillations at $\omega_{AF}$ are always present during the stick phase. Indeed, grains exhibit vibrational modes describing quasi-elliptic trajectories which are oriented, most of the time, parallel to the drive direction. In proximity of slip instabilities, conversely, the orientation of the ellipses changes and oscillations perpendicular to the drive direction emerge. These oscillations reduce the confining pressure and promote failure. The same mechanism is observed when the system undergoes an external perturbation at the frequency $\omega_{AF}$.    

\begin{figure*}[t!]
\begin{center}
\includegraphics[scale=0.32]{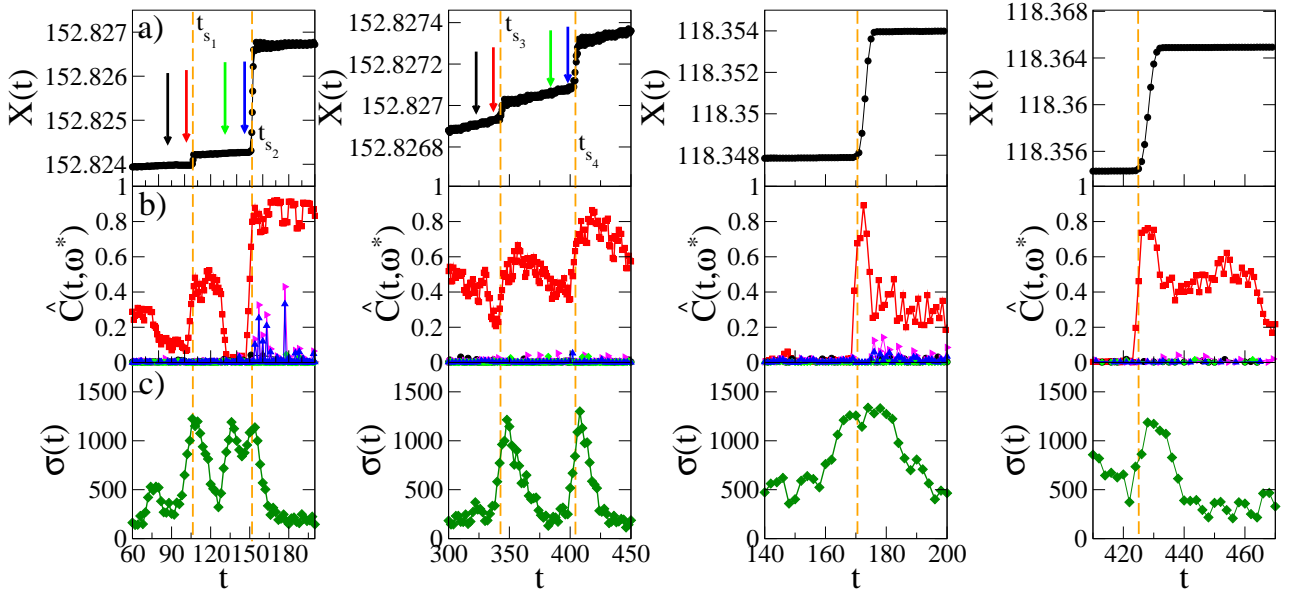} 
\end{center}
\caption{(Color online) 
 (Upper Panels) The top-plate position evolution along the direction of the applied drive ($x-$direction) exhibits stick-slip dynamics. The panels show four time intervals during which six slips of different sizes are observed at times 
indicated by the dashed vertical orange lines. 
The colored arrows show the times considered for the evaluation of  
$P(\theta)$ (Fig. \ref{fig4}).
(Central Panels) The power spectral density $\hat C(t,\omega)$ of the  autocorrelation function  $C(t,t')$ as function of the time $t$ for different values of $j\cdot d\omega$, with $d\omega=0.05 \pi$ and $j=20$ (blu),  $28$ (red), $36$ (purple). Values of $\hat C(t,\omega)$ significantly different than $0$ are only observed for $j=28$ corresponding to $\omega=\omega^{*}$.
(Lower Panels) The standard deviation of the angle $\theta$ formed by the particle
velocities with the $z$ axis as function of the time $t$. 
\label{fig1}
 }  
\end{figure*}

{\it The model}
The system is composed by $N$ spheres enclosed between two rigid rough plates of dimension $L_x\times L_y = 20d \times 5d$. Each plate is made of $L_x L_y /d^2$ spheres
of diameter $d$ placed in random positions in the fault plane, namely i.e. the $x-y$ plane. Spheres are shifted by a random $\delta z \in [0, d/2]$ in the  $z$-direction. In order to make the plates rigid, the particles keep their relative positions. This
preparation protocol ensures the roughness of both confining plates. The fault width is roughly of size $L_z \simeq 10d$.
While the bottom plate is kept fixed, the top one is subject to a constant pressure $p$ and attached
to a spring of elastic constant $k_m$ which is pulled at constant velocity $V$  along the $x$-direction. We employ a contact force model that captures the major features of granular interactions,
known as spring-dashpot model which also takes into account the presence of static friction~\cite{silbert,prl2010,giaccoscirep2014}.
The normal interaction between two contacting spheres is characterized by a spring constant $k_n = 2\cdot 10^{3} k_m$ and a damping coefficient $\gamma_n=50 \sqrt{k_{m}/m}$. 
Model parameters are chosen 
according to ref.~\cite{prl2010,epl2011} in order to have long stick phases 
interrupted by rapid plate displacements, i.e. the slips. 
The duration of the stick phase is inversely proportional to the driving velocity  $V$.
 We measure the mass in units of $m$, the length in units of $d$ and time in units of $\sqrt{m/k_{m}}$, the typical duration of a slip instability is of the order of one time unit. The confining pressure is $p=k_{m}/d$ and the driving velocity $V=0.01\, d/\sqrt{m/k_{m}}$.

In Fig.\ref{fig1} (upper panels) we plot the top-plate position in four time  intervals of different duration which present six slips. 
We evaluate the autocorrelation function of particle velocities
\begin{equation}
 C(t,t')= \frac{\sum_{i=1}^{N} \vec v_{i}(t)\cdot\vec 
v_{i}(t')}{\sum_{i=1}^{N} \vec v_{i}(t)\cdot \vec v_{i}(t)} ,
\end{equation}
where $\vec v_{i}$ is the velocity of the $i$-th particle.
More precisely, at each time $t$, we create a replica of the system 
decoupled from the external drive ($V=0$) and measure the particle velocity $\vec v_i(t')$ at subsequent times $t' \ge t$.
We  study  the temporal evolution of the power spectral density 
$\hat C(t,\omega)$
defined as the Fourier Transform, respect to $t'$, of $C(t,t')$.
We find that for all values of $\omega$,  the spectral density takes very small values so that  
$\hat C(t,\omega) < 0.01$ whereas much larger values are observed in a narrow range $\omega \in (1.3,1.5)\pi$ (red squares in Fig.\ref{fig1} central panels). 
In particular, Fig.\ref{fig1} (central panels) shows that  
$\hat C(t,\omega^{*})$, with  $\omega^{*}=1.4\pi$,  presents a non monotonic behavior as function of $t$. More precisely, 
$\hat C(t,\omega^{*})$ rapidly increases as $t$ approaches $t_s$, decreasing after $t_s$.
Since the dissipation is relevant only on time scales much larger than $1/\omega^*$, the integral of $\hat C(t,\omega)$  over the entire frequency range is always close to $1$. Hence,  the value of $\hat{C}(t,\omega^*)$ can be interpreted as the percentage of energy of the modes in the frequency range ($\omega^{*}-d\omega/2,\omega^{*}+d\omega/2$). 
The frequency $\omega^{*}$ can be related to the characteristic frequency $\omega_{AF}$ predicted by the AF scenario. Indeed, according to AF, 
the resonant frequency $\omega_{AF}$ characterizes the typical acoustic 
waves bouncing back-and-forth within the medium~\cite{giaccoprl2015}.  
 These waves propagate with velocity 
$v_a=\sqrt{M/\rho}$, where $M$
is the $P$-wave modulus and $\rho$ is the system density. 
The evaluation of $M$ in confined granular medium is very complicated and, indeed, experimental and numerical studies \cite{goddard90,makse2004,reichhardtPRE2015} indicate that it increases when the confining pressure is increased.
In a first approximation we can use the result $M\simeq k_{n}/(6d)$ obtained for
a single grain under  normal compressional stresses \footnote{In our model, a single grain under normal compression is deformed as a cube.  A compressional stress $\sigma_{ii}$ applied in the $i$-th direction, on the two faces perpendicular to the $i$-th direction, produces a deformation $2 \delta x_i$, along the $i$-th direction, with  $ k_n \delta x_i= \sigma_{ii} d^2$.
As a consequence 
$ \delta V/V \simeq-2 \sum_{i=1}^3
\delta x_i/d=-2/(d k_n) \sum_{i=1}^3 \sigma_{ii}= 
-6/(d k_n) P $,
where $P$ is the applied pressure}. Then  
using $\rho \simeq m N/(L_x L_y L_z)$ and taking into account   
that a time $T_a=2L_z/v_a$ is necessary  
to reach the bottom plate and return to the top,
the typical AF resonant frequency is
\be
\omega_{AF}= 2\pi/T_a = (\pi/L_z)\sqrt{k_n/(6 d \rho)}.
\label{oaf}
\ee 
This dependence of $\omega_{AF}$ on $k_n$ and $L_z$ has been explicitly verified  in ref. \cite{giaccoprl2015}. For the system under consideration here, 
we find from  Eq.(\ref{oaf}) $\omega_{AF} \simeq \omega^*=1.4\pi$,  
 given the values of the parameters of our simulations. This indicates that vibrational modes at the characteristic frequency $\omega_{AF}$  appear at the onset of each slip.  Experiments on confined granular materials \cite{johnsonjia2005} show that the wave speed depends on the applied external pressure, suggesting the existence of a critical frequency below which acoustic emissions
cannot be activated. In our simulations grains are assumed to deform only elastically and we therefore expect that, in our system, the AF mechanism can be activated at  all pressures.


\begin{figure}[!t]
\begin{center}
\includegraphics[scale=0.52]{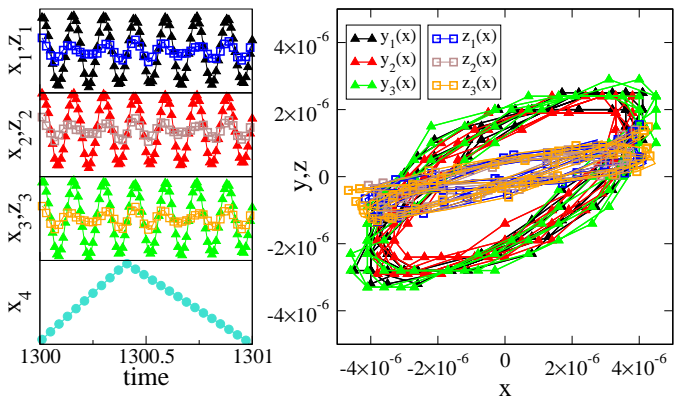}
\;\;\;\;\;\;\;\;\;\;\newline
\vspace{0.1cm}
\includegraphics[scale=0.51]{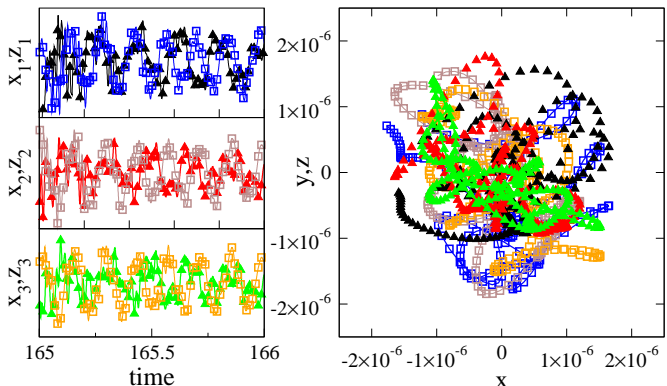}
\end{center}
\vskip -0.76cm
\caption{(Color online) (Upper left panels) Time dependence of the $x$ (filled triangles) and $z$-coordinate (open squares) of the position of four nearest neighbor particles at the time $t_{s_1}-t =20$, far from the slip. The vertical scale of $x_4$ is $3000$ times larger than the scale of $x_1,x_2,x_3$. (Upper Right panels)  The position components  $y_1,y_2,y_3$  and $z_1,z_2,z_3$ are plotted as a function of  $x_1,x_2,x_3$, respectively, at the time $t_{s_1}-t =20$. Each trajectory has been shifted to be all  centered  in $(0,0)$. The same symbols and colors as in the left panels. (Lower panels) The same as upper panels at the time $t_{s_1}-t =5$, at the onset of the slip.}
\label{fig2}
\end{figure}


\begin{figure}
\begin{center}
\includegraphics[scale=0.52]{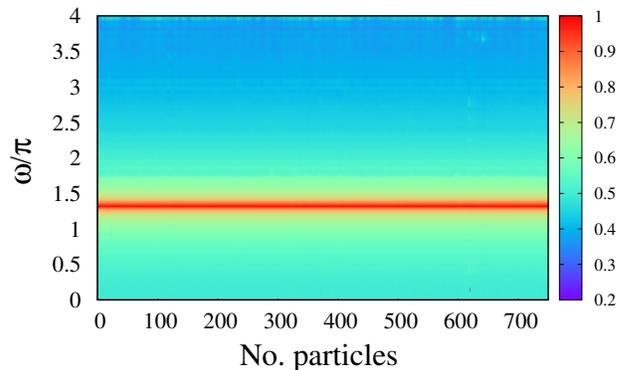}
\end{center}
\vskip -0.76cm
\caption{(Color online) 
The map of the power spectral density, $\hat C_{x_i}(\omega)$, of particles position  as function of $\omega$, for all particles. The horizontal axis represents the particle index. The intensity of the power spectrum can be obtained from the color code.}  
\label{fig3}
\end{figure}

To understand the mechanisms responsible for the increase in $\hat C(t,\omega)$ for the frequency $\omega_{AF}$ at the onset of each slip, we follow the trajectories of each grain inside the stick phase. We find that, because of the high granular density, the large majority of particles are always in contact with their neighbors forming an almost rigid structure, i.e. the backbone. Conversely, a small fraction (less than $10\%$) of particles, the rattlers,
 are located inside the cages formed by the particles in the backbone \cite{torquatorevmod2010,atkinson2016} and most of the time do not interact with  other particles \cite{energyrelax2017}. 
These cages form soon after each slip and keep substantially their configuration unaltered during the whole stick phase, as shown in Fig. 1 in the Supplementary Material \cite{suppmat}. Rattlers present a kinetic energy order of magnitude larger than the backbone energy \cite{energyrelax2017}. They are then identified with the shear transforming zone \cite{lieou2015,lieoujgr2017} according to a local strain measure as $D_{min}^2$ introduced in ref.\cite{falklanger98}. Nevertheless, rattlers are not directly responsible of slip instabilities caused by the collapse of the force chains made by backbone particles. 
In the left panels of 
Fig. \ref{fig2} we  plot the $x$-position of four  neighboring particles during a 
short time window: Three particles $(x_1,x_2,x_3)$
 exhibit regular oscillations along the $x$-direction. 
Differently, the particle $4$ is a rattler and moves along a straight line up
to an abrupt change in direction caused by a collision. 
In the following we restrict the study to backbone particles. As shown  
in the left panel of Fig.\ref{fig2}, the particle motion in the $x$-direction corresponds to vibrational modes at 
a characteristic frequency $\omega=1.4\pi \simeq \omega_{AF}$. 
More precisely, when we superimpose the centers of each 
trajectory in a common point, as  in Fig.\ref{fig2} (upper right panel), trajectories are roughly confined in a plane and exhibit an elliptic-like shape.
The same oscillating behavior  (not shown) is observed for the other backbone particles and for all time intervals  during the stick phase.
This pattern is recovered for all particles as confirmed by the power spectrum of the particle position $\hat C_{x_i}(\omega)$ obtained from the Fourier Transform of the $i$-th particle position $x_i(t)$. Fig. \ref{fig3}, indeed, shows that all particles present characteristic oscillations at a frequency $\omega \simeq \omega _{AF}$. 
 Hence, this study shows that vibrational modes at the characteristic frequency $\omega_{AF}$ do not form at the onset of slips but are already present inside the system at all times. The energy responsible for these oscillations originates from the energy stored, during the stick phase,   through the spring which couples the system to the external drive. 
   Most of this energy is released very rapidly during the slip but a significant fraction contributes to the activation of harmonic oscillations. Because of the vertical confinement only the mode at the frequency $\omega_{AF}$ (Eq.(\ref{oaf})) survives. 
Even if these modes have been explained in terms of compressional waves propagating along the $z-$direction, because of the heterogeneous structure of the granular packing, these waves induce also displacements along the $x-$ and $y-$ directions. Far from the slip, the confinement along the $z-$direction and periodic boundary conditions along $x$ and $y$, lead to $x−$ and $y−$ displacements
larger than $z$-displacements.

\begin{figure}[!t]
\begin{center}
\includegraphics[scale=0.33]{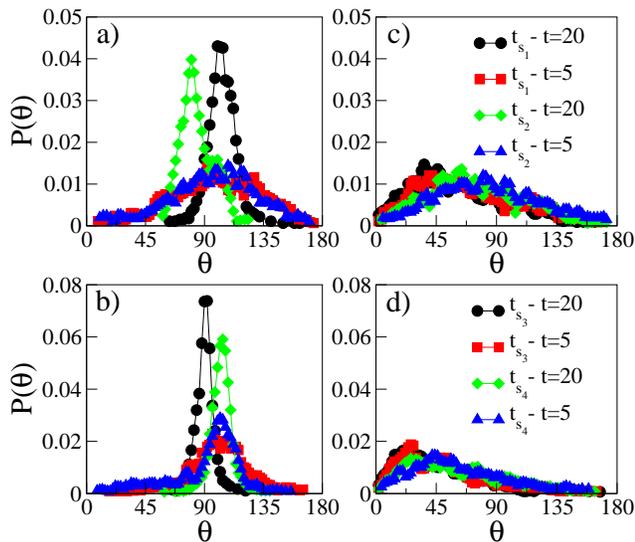}
\end{center}
\vskip -0.76cm
\caption{(Color online) (Left panels) The distribution of the angle $\theta$ formed by the particle velocities with the $z$-axis in the unperturbed evolution. Each color corresponds to a different time $t$ indicated by the vertical arrows in the upper panels of Fig. 1. In the upper panel we consider the temporal interval $t \in [60,190]$ and in the lower panel the temporal interval $t\in [300,450]]$.
(Right Panels) The quantity $P(\theta)$ under a pressure perturbation evaluated at the same times, identified by the same color codes,  of  left panels. }  
\label{fig4}
\end{figure}

We wish to stress that $\hat C_{x_i}(\omega)$ is the Fourier Transform of a one-time quantity $x(t)$ whereas $\hat C(t,\omega)$ is obtained from the correlation function $C(t,t')$ which is a two-time quantity. Hence the existence of a temporal interval with $\hat C(t,\omega_{AF}) \simeq 0$ before $t_s$  indicates a decorrelation of grain velocities.
To rationalize the origin of this decorrelation we investigate the orientation of the
quasi-elliptic trajectories  during the evolution. We find that ellipses are preferentially oriented along the $x-$direction and that, just before the occurrence of a slip, oscillations along the $z-$direction 
appear. This is confirmed by the distribution $P(\theta)$  of the angle $\theta$ formed 
by the particle velocities with  the $z$ axis. In Fig. \ref{fig4}  we plot  $P(\theta)$ evaluated at different times before and after $t_s$. We find that, for all slips, at large temporal distances from $t_s$, $P(\theta)$ is sharply peaked at $\theta \simeq 90^o $, corresponding to an oscillatory motion in the $x-y$ plane. This distribution does not change significantly during the evolution and  only  in proximity of the slip time 
 it spreads towards smaller values of $\theta$ (red squares and  blue triangles). Therefore, when $t$ approaches $t_s$ we find the presence of oscillations also in the direction  parallel to the $z$-axis ($\theta \simeq 0$). This is confirmed by the behavior of the $z$-coordinates as function of time and as function the $x$-coordinate (Fig.\ref{fig2}). Far from the slip (upper panels of Fig.\ref{fig2}), the displacement in the $z$-direction presents oscillation at the frequency $\omega_{AF}$. As already observed $z-$displacements
are small compared to the $x-$displacements and the trajectory is mostly confined in the $x$-$y$ plane ($\theta=90^o$) (upper left panels of Fig.\ref{fig2}). At the onset of slip instability   (lower panels of Fig.\ref{fig2}) the angle $\theta$ is no further stable and $z-$ displacements, of size comparable to $x-$displacements, are indeed observed. This behavior is also  
clearly enlightened by animations presented in the  Supplementary Material \cite{suppmat}. The above findings support the hypothesis of 
weakening by AF. Indeed when oscillations are confined in the $x$-$y$ plane 
($\theta \simeq 90^o$) they do not affect the confining pressure. Conversely, when $\theta \simeq 0 ^o$, oscillations can reduce the confining pressure promoting failure. 
In order to support this interpretation in Fig. \ref{fig1} (lower panels) we plot the standard deviation of the the $\theta$ angle 
$$\sigma(t) = \frac{1}{N} \sum_{i=1}^N \theta_i(t)^2- \left( \frac{1}{N} \sum_{i=1}^N \theta_i(t) \right )^2 $$
where $\theta_i(t)$ is the angle formed by the velocity of the $i$-th particle at the time $t$ with the $z$-axis. We find small values of $\sigma(t)$ when most of the trajectories are aligned in the $x-y$ plane ( $\theta \simeq 90^o$) whereas larger values are found when oscillations along the $z-$direction ($\theta \simeq 0 ^o$) are observed. Fig. \ref{fig1} (lower panels) shows that $\sigma(t)$ typically increases before slip instabilities  where $z-$oscillations increase the slip probability. A decay of $\sigma(t)$ is conversely observed at times $t>t_s$. Interestingly we observe that the onset of the increase of $\sigma(t)$ slightly anticipates the onset of the increase of $\hat C(t,\omega_{AF})$. This indicates that the configuration with all ellipses oriented along the $x-$direction is no longer stable since oscillations along other directions start to appear in the system.  
Finally the increase of $C(t,\omega_{AF})$ immediately before $t_s$ indicates that a coherent behavior of grain trajectories is recovered at the onset of slip instability with $z-$oscillations  promoting failure. These oscillations along the $z-$direction are probably activated by collisions of rattlers with backbone particles, sufficiently energetic to destabilize  oscillations originally confined in the $x-y$ plane. The occurrence time of these collisions appear to be non-predictable.

The overall picture is confirmed by the evolution of the quasi-elliptic trajectories when a  compressive periodic perturbation is applied at the resonant frequency $\omega_{AF}$, which increases the confining pressure of  $5 \%$.
 The response to the external perturbation clearly depends on the temporal distance from the slip instability. When the system is far from the slip, the external perturbation promotes a change in the ellipse orientation which tends to align along the $z-$direction. 
This is confirmed by animation (Supp. Mat. \cite{suppmat}) and by the distribution of $\theta$ (Fig. \ref{fig4} right panels) which, in presence of a perturbation,  moves towards smaller values of $\theta$. The same behavior is observed at all times during the stick phase.
Conversely, only close to  instabilities the trajectories are 
weakly affected by the perturbation and indeed  only small differences are found in the angle distribution  $P(\theta)$ with or without the perturbation. 

Summarizing, our results provide support to the AF scenario
indicating that vibrational modes at the characteristic frequency $\omega_{AF}$ are present during the entire stick phase. These vibrations affect the confining pressure only at the onset of slip instabilities, or in presence of external perturbations, when oscillations along the $z-$direction are observed.

\end{document}